\documentclass[fleqn,usenatbib]{mnras}

\usepackage{mathtools, cuted}
\usepackage{comment}

\usepackage[T1]{fontenc}

\DeclareRobustCommand{\VAN}[3]{#2}
\let\VANthebibliography\thebibliography
\def\thebibliography{\DeclareRobustCommand{\VAN}[3]{##3}\VANthebibliography}

\usepackage{subcaption}
\usepackage{graphicx}	
\usepackage{amsmath}	
\usepackage{amssymb}	
\usepackage{newtxtext,newtxmath}

\title[Disintegrating plug]{Plug Disintegration in GRB Jet Eruption}

\author[A. Yalinewich and P. Beniamini]{Almog Yalinewich$^{1}$\thanks{E-mail: almog.yalin@gmail.com} and
Paz Beniamini$^{2,3}$
\\
$^{1}$Canadian Institute for Theoretical Astrophysics, 60 St. George St., Toronto, ON M5S 3H8, Canada\\
$^{2}$ Department of Natural Sciences, Open University of Israel, 1 University Road, 43107 Ra’anana, Israel\\
$^{3}$ Astrophysics Research Center of the Open University (ARCO), The Open University of Israel, P.O Box 808, Ra’anana 43537, Israel
\\
}

\date{Accepted XXX. Received YYY; in original form ZZZ}

\pubyear{2020}

\begin{document}
\label{firstpage}
\pagerange{\pageref{firstpage}--\pageref{lastpage}}
\maketitle

\begin{abstract}
In this work we consider the eruption of a tenuous relativistic hydrodynamic jet from a dense baryonic envelope. As the jet moves out and away, it carries along and continues to accelerate a layer of baryonic material which we refer to as the plug. We solve the relativistic equations of motion for the trajectory of the plug, and verify it using a relativistic hydrodynamic simulation. We show that under these conditions, the plug breaks up at a radius larger by a factor of a few from the radius of the envelope, due to the onset of the Rayleigh Taylor instability. After breakup the jet continues to accelerate to higher Lorentz factors while the plug fragments maintain a moderate Lorentz factor. The presence of slower moving ejecta can explain late time features of GRBs such as X ray flares without recourse to a long lived engine.
\end{abstract}




\section{Introduction}

Gamma ray bursts (GRBs) are intense, short, extragalactic flashes of gamma rays \citep[see][and references therein for a review]{Piran2004TheBursts, Levan2016Gamma-RayProgenitors}. There are two types of gamma ray bursts: short gamma ray bursts, which typically last a fraction of a second, and long gamma ray bursts which typically last tens to hundreds of seconds. The two kinds occur through two different channels: the short ones have been associated with merging neutron stars \citep[e.g.][and references therein]{Nakar2019TheMergers}, and the long ones with stripped envelope core collapse supernovae \citep{Kumar2015TheJets}. 

In both cases, a relativistic tenuous jet breaks out of a dense baryonic envelope. The most popular model for long gamma ray bursts is the collpsar scenario \citep[][and references therein]{Hartmann2000HypernovaeBursts}. This scenario begins when the core of a massive star exhausts its nuclear fuel and begins to collapse. The core collapses to form a relativistic compact object (either a black hole or a neutron star) and begins to accrete the stellar material. The compact object launches a relativistic jet that burrows its way out of the star.

Short gamma ray bursts occur when neutron stars merge \citep[see][and references within]{Berger2014Short-durationBursts}. As a result of the collision, ejecta is expelled primarily in the direction normal to the plane of motion of the two neutron stars (though some ejecta is also expelled in the plane of motion due to tidal tails). Next, the two neutron stars merge to become a fast spinning compact object. This object could either be a stable neutron star, a black hole, or a hyper or supra massive neutron star that later collapses into a black hole when it spins down or accretes more material, though a long lived neutron star is disfavoured \citep{Beniamini2021SurvivalMergers}. The merger product does not absorb all the material of its parents immediately, so some of it forms an accretion disc around it. The merger product accretes this material at a super Eddington rate and launches a jet. This jet collides with, and eventually erupts out of a baryonic envelope made up of a mixture of previously expelled ejecta and disc wind. This situation is very similar to jet eruption in long gamma ray bursts. The main differences between the two scenarios is that in the case of short gamma ray bursts the mass of the baryonic envelope is smaller, the jet engine is active for a shorter time, and the density distribution is steeper \citep{Kathirgamaraju2019ObservableAfterglow, Hajela2019TwoEjecta, Balasubramanian2021ContinuedPost-merger}. The outward motion of the ejecta is not important, because the jet is moving much faster, and so we can consider the ejecta to be stationary. We note that this condition is satisfied only after the energy emitted by the jet exceeds the kinetic energy of the upstream material \citep{Duffell2018JetJet, Beniamini2020ReadyFormation} and not from the moment the jet is launched.

One of the open questions in the study of gamma ray bursts is how much baryonic material gets picked up by the jet. The formation and breakup of this hypothetical layer ahead of the jet,sometimes also called ``plug" or ``cork'', has been studied using numerical simulations. \cite{ Zhang2004TheBursts, Lopez-Camara2012Three-dimensionalStars, Mizuta2013OpeningJETS} ran two and three dimensional simulations of hydrodynamic jet eruptions from stellar envelopes. The plug persevered only in the two dimensional simulations. In three dimensions, the plug quickly disintegrated and the jet material flowed around the clumps, following the path of least resistance. \cite{Gottlieb2020TheJets} showed that the instability that destroys the plug in 3D simulations can be suppressed by subdominant toroidal magnetic fields, so that plug survives when magnetic fields are present.

It has been suggested that the existence and survival of this plug can explain some observational features in gamma ray bursts \citep{Eichler2014CloakedBursts}. \cite{Eichler2004AnBursts} showed that the Amati relations, connecting the isotropic equivalent prompt gamma-ray energy and its spectral peak, can be explained by an oblique viewing angle of a jet obscured by a spherical cap. \cite{Waxman2003CollapsarBursts} proposed that the plug is shocked multiple times after it bursts from the stellar envelope, and that each such shock can give rise to a hard X-ray pulse. \cite{Eichler2007ABursts, Eichler2008SpectralScatterers} showed that Compton scattering from radiatively accelerated baryons can explain subpulse structure, as well as spectral evolution observed in some gamma ray bursts. \cite{Vyas2020ABursts} explored this idea further by running Monte Carlo simulations of Compton scattering of photons from the relativistically moving shell, and found that this process can explain the spectral slopes of the emitted radiation. \cite{Duffell2015FromAfterglows} studied how the accumulation of baryonic material ahead of the jet slows it down and delays the dissipation of energy. They suggested that the formation of a plug can thus explain late time features like the gamma ray plateau. It has also been suggested that a baryonic shell ahead of the jet material could thermalise the photons from the central engine, and that this mechanism could explain the narrowness of the prompt spectrum \citep{Thompson2006DecelerationBursts, Thompson2014PulseBaryons}.

One of the scenarios we consider in this work is a plug that is accelerated by the jet material up to a certain point where it decouples from the jet. After decoupling, the jet can continue to accelerate to attain a higher Lorentz factor. It has been proposed \cite{Beniamini2016X-rayOrigin} that a slower moving shell of material can explain narrow X ray flares (erratic X-ray brightening episodes often seen in gamma ray burst afterglows, \citealt{Burrows2005BrightAfterglows, Falcone2007TheEnergetics,  Chincarini2010UnveilingBursts, Margutti2010Lag-luminosityEmission}) observed in the afterglows of some GRBs. As outlined in \cite{Beniamini2016X-rayOrigin}, the advantages of ejecting the flare producing material together with the prompt producing jet, are that it: (i) can tap into a larger energy reservoir, available at the start of the burst and (ii) enables the slower and less energetic material to pass through the cavity created by the main jet and erupt from the surrounding material with little expenditure of energy and with a narrow opening angle (relative to the inverse of its Lorentz factor) allowing for very steep decays at the end of the flares, as observed. Furthermore, depending on the physics of the energy dissipation, the ratio of flare to prompt time-scales can be a strong function of the ratio of their respective Lorentz factors, thus naturally explaining the wide diversity in flare occurrence times.
A disintegrated plug may naturally posses the right luminosity and Lorentz factor needed to account for observed X-ray flares.

The plan of the paper is as follows. In section \ref{sec:jet_prop} we calculate the properties of the baryonic mass entrained by jet as a function of the jet's properties. In section \ref{sec:plug_acceleration} we describe the mathematical model for the one dimensional acceleration of the plug by the jet, and verification of these calculation using a one dimensional Lagrangian relativistic hydrodynamic simulation. In section \ref{sec:plug_dissolution} we discuss the condition that determines the point at which the plug will break up due to the onset of the Rayleigh Taylor instability or become transparent. In section \ref{sec:conclusions} we discuss the relations of our results to previous works and implications for gamma ray bursts.

\section{Jet Propagation and Eruption} \label{sec:jet_prop}

For simplicity, we  consider a central engine that emits only photons with some constant luminosity $L$ and at some small opening angle $\alpha \ll 1$. Ignoring lepton and baryon contamination is justified as long as their rest frame density is much lower than the upstream density of the envelope material, a reasonable assumption at least until the jet erupts from the surface of the envelope. We also assume a constant density $\rho_c$ in the vicinity of the engine. Close to the jet engine, the jet material expands ballistically, but at some distance the flow becomes collimated. Inside the collimated jet the Lorentz factor of the jet remains roughly constant $\gamma \approx 1/\alpha$ \citep{Bromberg2011TheMedia}.

As the jet moves outward it encounters a declining density profile. As the density declines, collimation weakens, until at some finite distance the flow decollimates, and beyond that point the flow expands at an angle $\alpha$, while continuing to travel with a bulk Lorentz factor of order $1/\alpha$. Therefore, all the material ahead of the shock cannot move in the lateral direction and gets entrained by the jet. Due to the difference in the envelope density distributions, we treat long and short gamma ray bursts separately.

\subsection{Long Gamma Ray Bursts}

In the case of long gamma ray bursts, the density of the envelope is determined by hydrostatic equilibrium. Therefore, close to the star's surface, the density can be described by a polytropic profile
\begin{equation}
    \rho_a \approx \frac{M_e}{R_e^3} \left(\frac{x}{R_e}\right)^{\omega}
\end{equation}
where $M_e$ is the mass of the progenitor star, $R_e$ is its radius and $x$ is the depth measured from the stellar surface. Decollimation happens when the upstream density cannot considerably decelerate the jet, so it expands with a Lorentz factor $1/\alpha$. Pressure balance between the shocked jet and shocked envelope material at the moment of decollimation yields
\begin{equation}
    \frac{L}{\alpha^2 x^2 c} \approx \frac{\rho_a c^2}{\alpha^4}
\end{equation}
where $\alpha$ on the left hand side comes from geometry, and on the right hand side it enters through the Lorentz factor. By solving for $x$ we can obtain the eruption depth $x_{er}$
\begin{equation}
    x_{er} \approx L^{\frac{1}{\omega + 2}} M_{e}^{- \frac{1}{\omega + 2}} R_{e}^{\frac{\omega + 3}{\omega + 2}} \alpha^{\frac{2}{\omega + 2}} c^{- \frac{3}{\omega + 2}} \, .
\end{equation}
If we assume a radiative envelope $\omega=3$, we can evaluate the expression for the eruption depth
\begin{equation}
    x_{er} \approx 0.16 \tilde{L}^{0.2} \tilde{R_{e}}^{1.2} \tilde{\alpha}^{0.4} \tilde{M_{e}}^{-0.2} \,  R_{\odot}
\end{equation}
where $\tilde{L}=L/10^{50} \, \rm {erg/s}$, $\tilde{R}_e = R_e/3 R_{\odot}$, $\tilde{\alpha} = \alpha/0.1$ and $\tilde{M}_e = M_e / 10 M_{\odot}$. The mass carried by the jet is given by
\begin{equation}
    m_{er} \approx L^{\frac{\omega + 3}{\omega + 2}} M_{e}^{- \frac{1}{\omega + 2}} R_{e}^{\frac{- \omega \left(\omega + 2\right) + \omega \left(\omega + 3\right) + 3}{\omega + 2}} \alpha^{\frac{2 \left(2 \omega + 5\right)}{\omega + 2}} c^{- \frac{3 \left(\omega + 3\right)}{\omega + 2}} \, .
\end{equation}
For our fiducial parameters we can estimate this mass
\begin{equation}
    m_{er} \approx 2.0 \cdot 10^{-9}  \tilde{L}^{1.2} \tilde{R_{e}}^{1.2} \tilde{\alpha}^{4.4} \tilde{M_{e}}^{-0.2} \, M_{\odot}.
\end{equation}

In the discussion above we neglected the stellar wind, which could be important since Wolf Rayet stars could be GRB progentiros \citep{Thompson2006DecelerationBursts}. A dense enough stellar wind can keep the jet collimated longer, so that eruption happens from a larger radius. Moreover, since the the gas is not in hydrostatic equilibrium, it does not terminate abruptly at a finite radius. Such an eruption is similar to the case of short gamma ray bursts, which will be analysed in the next section, where it will be shown that the mass of the entrained material is independent of the envelope mass, so the results should be the same both in the case of short gamma ray bursts and long gamma ray bursts with a wind. Finally, we want to mention the possibility of pre - acceleration of the wind to relativistic velocities due to interaction with radiation that leaks through the plug \citep{Thompson1999RelativisticBursts}. This effect can increase the mass of the entrained baryon rich material, but we will not consider it in this work.

\subsection{Short Gamma Ray Bursts}

In the case of short gamma ray bursts, the envelope is made up of material ejected from the merging neutron stars. Since this material is not in hydrostatic equilibrium, the density does not terminate abruptly at some finite radius, but rather declines in a steep, albeit smooth way. The mass - velocity distribution of the ejecta can be described by a Gaussian \citep{Radice2018BinaryNucleosynthesis}
\begin{equation}
    m \approx M_e \exp \left(-v^2/v_e^2\right)
\end{equation}
where $v_e \approx 0.2 c$. Using this mass velocity distribution and assuming a homologous expansion $v = r/t$, we can obtain the density distribution
\begin{equation}
    \rho_e \approx \frac{m}{r^3} \approx \frac{M_e}{r^3} \exp \left(-\frac{r^2}{t_e^2 v_e^2}\right)
\end{equation}
where $t_e \approx 0.1 \, \rm s$ is the time between the expulsion of the ejecta and jet launching \citep{Beniamini2020ReadyFormation}. Since $v_e$ and $t_e$ always appear together, we replace them with a single parameter $h_e = t_e v_e$. The condition for decollimation in this case is
\begin{equation}
    \frac{L}{\alpha^2 h_e^2 c} \approx \frac{\rho_e c^2}{\alpha^4} \, .
\end{equation}
This condition yields a transcendental equation for $r$, which we can solve by assuming that $r \approx h$, so we can replace all occurrences of $r$ in the equation by $h$, except for the exponent. The solution can therefore be approximated by
\begin{equation}
    r \approx h_e \sqrt{\ln \left(\frac{M_e c^3}{L \alpha^2 h_e}\right)}
\end{equation}
which, for our fiducial model, evaluates to
\begin{equation}
    r_{er} \approx 0.02 \tilde{h}_e \sqrt{ \ln \left(\frac{\tilde{M}_e}{\tilde{\alpha}^2 \tilde{L} \tilde{h}_e}\right)} \, R_{\odot}
\end{equation}
where $\tilde{M}_e = M_e/10^{-3} \, M_{\odot}$, $\tilde{h}_e = h_e/0.008 R_{\odot}$, $\tilde{L} = L/10^{50} \, \rm ers/s$ and $\tilde{\alpha} = \alpha/0.1$.

Unlike the previous case, after decollimation at a radius $r_{er}$, it will still have to travel a distance comparable to $r_{er}$ to accumulate most of the ejecta it will accelerate. The the swept up mass is therefore
\begin{equation}
    m_{er} \approx \rho_e h_e^3 \alpha^2 \approx \frac{\alpha^4 L h_e}{c^3} \, .
\end{equation}
For our fiducial parameters, we can evaluate this expression
\begin{equation}
    m_{er} \approx 10^{-10} \tilde{\alpha}^4 \tilde{L} \tilde{h}_e \, M_{\odot} .
\end{equation}

\section{Plug Acceleration} \label{sec:plug_acceleration}

\begin{figure*}
\begin{tabular}{ccc}
\subfloat{\includegraphics[width = 2.1in]{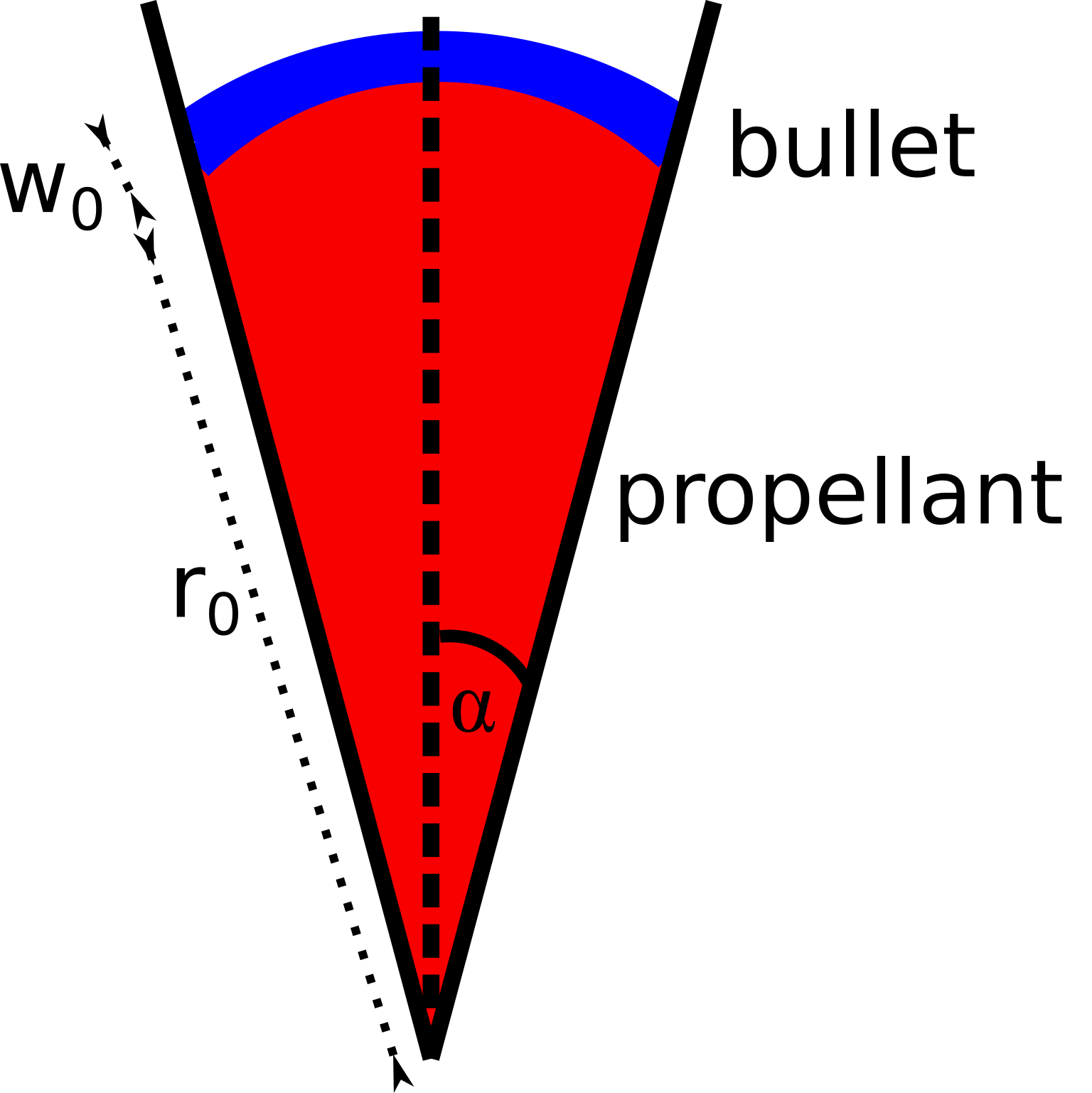}}&
\subfloat{\includegraphics[width = 2.1in]{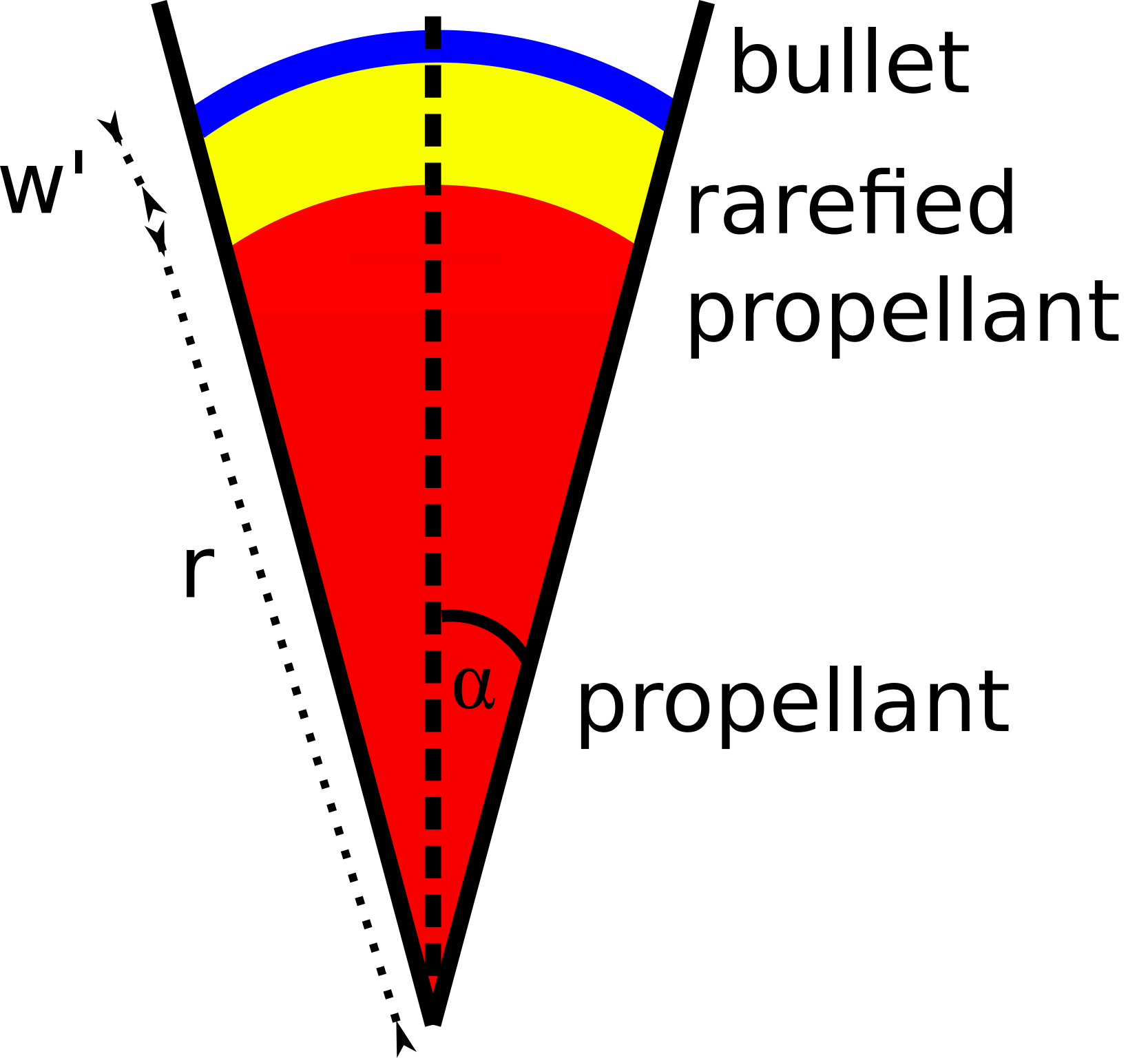}}&
\subfloat{\includegraphics[width = 2.1in]{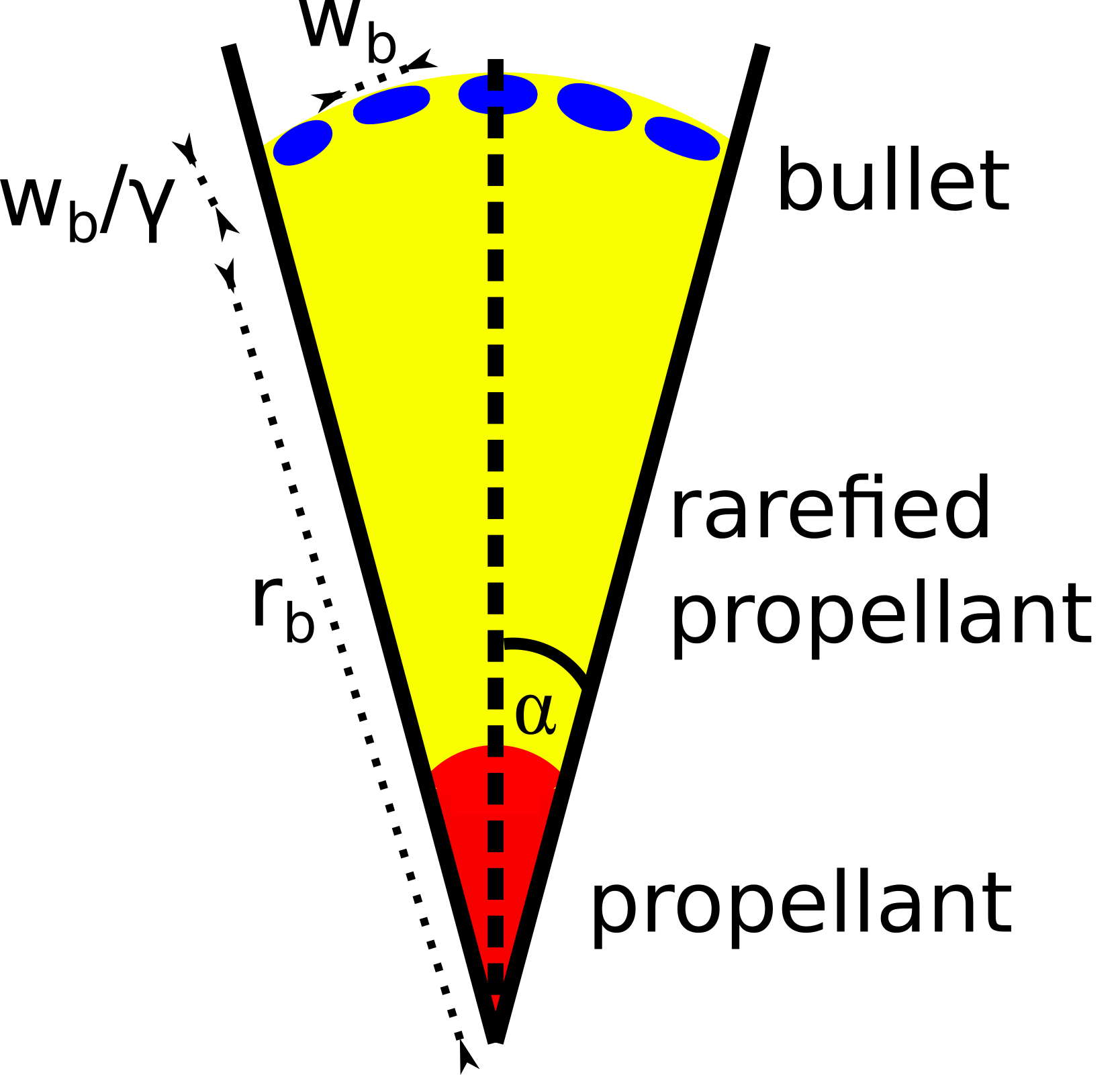}}
\end{tabular}
\caption{Schematic illustration of the initial conditions (left), plug acceleration (middle) and breakup (right). All illustrations are not to scale. Inside a conical barrel is a hot gas (red region) bounded by a denser plug (blue). We assume the plug always stretches to cover the propellant. As the plug moves forward, a rarefaction wave emerges from the plug - propellant interface and moves backward and into the propellant (marked in yellow in the illustration). When the plug reaches a certain critical radius, it breaks up due to the Rayleigh Taylor instability.}
\label{fig:schematic}
\end{figure*}

Let us consider an infinite and rigid cone with an opening angle $\alpha$. Inside the cone, at a distance $r_0$ from the apex is a spherical cap of dense material whose thickness is $w \ll r_0$, which we call the plug. The region $r<r_0-w$, which we refer to as the chamber, is filled with a hot gas, which we call the propellant. The initial conditions of the system, as well as subsequent stages of its evolution, are illustrated in figure \ref{fig:schematic}. Both the plug and the propellant are described by an ideal gas equation of state, but we allow the adiabatic index of the propellant $\eta$ to be different from that of the plug $\eta_b$. In this section we obtain the trajectory of the plug. To help explain this problem, we've included analyses of simpler versions of this problem in appendix \ref{sec:ped_rev}.

The ultra relativistic equation of motion for the plug is given by 
\begin{equation}
    c \frac{d}{d t} \left(M \gamma\right) = A p
\end{equation}
where $c$ is the speed of light, $M$ is the rest inertia of the plug, $A$ is the cross sectional area of the plug and $p$ is the propellant pressure. The effective mass $M$ is equal to the unshocked mass of the plug $m_{er}$ and the thermal energy. We are interested in the case where the thermal energy in the plug dominates over the rest mass energy density, and hence the inertia depends on pressure. We assume the plug expands isentropically, so the volume scales as $p^{1/\eta_b}$ and so the inertia scales as $p^{1-1/\eta_b}$.
As the plug moves faster, the pressure behind the plug decreases, such that at asymptotically long times the plug tends to a terminal velocity. 
To solve the equation of motion we need a relation between the velocity and pressure. Such a relation is given by the Riemann invariant. The ultra relativistic Riemann invariant in spherical geometry is
\begin{equation}
    J_+ = p \gamma^{\eta/\sqrt{\eta-1}} t^{2 \eta / \left(1+\sqrt{\eta-1}\right)} \, .
\end{equation}
The original derivation for the spherical Riemann invariant for $\eta=4/3$ appears in \cite{Oren2009DiscreteExplosions}, and we also include a complete derivation for a general $\eta$ in appendix \ref{sec:spheric_ri}. The flow is always assumed to be in the radial direction at a speed close to the speed of light, so $t =r/c$ where $r$ is the radius. The pressure is therefore given by
\begin{equation}
    p = p_1 \left(\frac{\gamma}{\gamma_1}\right)^{-\frac{\eta}{\sqrt{\eta-1}}} \left(\frac{t}{t_1}\right)^{-\frac{2 \eta}{1+\sqrt{\eta-1}}}
\end{equation}
where $p_1$ and $\gamma_1$ are the initial pressure and Lorentz factor at the initial time $t_1=r_0/c$. We also assume that the jet has an opening angle larger than the beaming angle $\alpha > 1/\gamma$ so that the effective cross section is $A = \alpha^2 t^2 c^2$. The solution to the equation of motion at late times $t \gg M c/p_0 \alpha^2 c^2 t_1^2$ is given by
\begin{equation}
    \gamma \approx \gamma_1 \left(\frac{\alpha^2 t_1^3 c^2 p_1}{\gamma_1 M c}\right)^{\frac{\sqrt{\eta - 1}}{\sqrt{\eta - 1} + 1}} \left(\frac{t}{t_1}\right)^{\frac{\sqrt{\eta - 1} \left(3 \sqrt{\eta - 1} + 1\right)}{\left(\sqrt{\eta - 1} + 1\right)^{2}}} \, .
\end{equation}
The acceleration at late times is given asymptotically by
\begin{equation}
\dot{\gamma} \approx \frac{\gamma_1}{t_1} \left(\frac{\alpha^2 t_1^3 c^2 p_1}{\gamma_1 M c}\right)^{\frac{\sqrt{\eta - 1}}{\sqrt{\eta - 1} + 1}} \left(\frac{t}{t_1}\right)^{\frac{\sqrt{\eta - 1} \left(3 \sqrt{\eta - 1} + 1\right)}{\left(\sqrt{\eta - 1} + 1\right)^{2}}-1} \, .
\end{equation}
This acceleration continues until the shell attains its terminal Lorentz factor \citep{Yalinewich2017AnalyticBreakout}. This happens when the fluid exhausts its thermal energy, so the rest frame energy is comparable to the rest mass energy density, i.e. $p \approx \rho c^2$, where $\rho$ is the rest frame baryon mass density. The pressure as a function of time is given by
\begin{equation}
    p = p_1 \left(\frac{\alpha^2 t_i^3 c^2 p_1}{\gamma_1 M c}\right)^{\psi_1} \left(\frac{t}{t_i}\right)^{\psi_2}
\end{equation}
where $\psi_1 = - \frac{\eta}{\sqrt{\eta - 1} + 1}$ and $\psi_2 =- \frac{\eta \left(5 \sqrt{\eta - 1} + 3\right)}{\left(\sqrt{\eta - 1} + 1\right)^{2}}$. For $\eta=4/3$, $\psi_1 = -0.85$ and $\psi_2 = -3.2$.

\begin{figure*}
    \centering
    \includegraphics[width=0.9\textwidth]{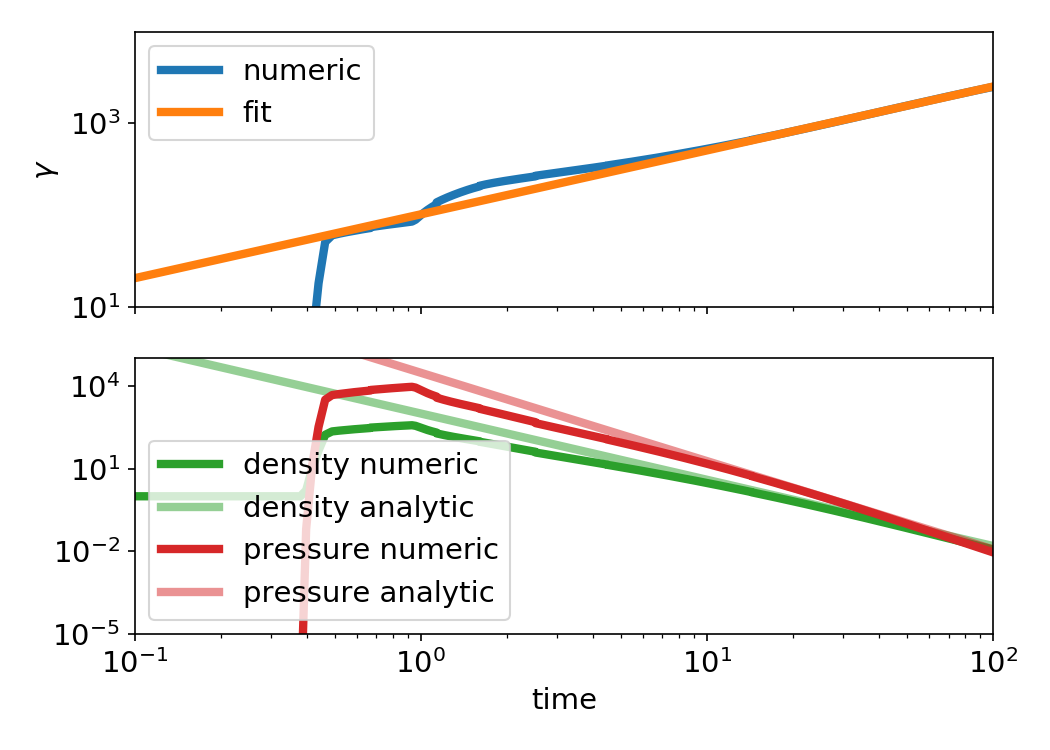}
    \caption{Histories of hydrodynamic values of a single fluid element inside the plug, in the case of an acceleration inside a spherical barrel. The numerical slope of the Lorentz factor history is $d \ln \gamma / d \ln t = 0.69$, which is very close to the theoretical value 0.63. The bottom panel shows the evolution of the pressure and density. The thermal pressure initially exceeds the rest mass energy density, but as the plug cools and expands the pressure eventually drops below the rest mass energy density.}
    \label{fig:spherical_simulation}
\end{figure*}

To verify our theoretical predictions we ran a one dimensional, spherically symmetric, Lagrangian, special relativistic numerical simulation. The initial computational domain lay between $0 < r < 10$, where $0<r<9$ represents the chamber and $9<r<10$ represents the plug, both of which are described by an ideal gas equation of state with an adiabatic index $\eta=4/3$. Inside the chamber the density was low ($10^{-8}$) and the pressure high ($10^{10}$) while in the plug the density was high (1) and the pressure low ($10^{-8}$). The velocity was zero everywhere, and the speed of light in simulation units is 1. We divided the computational domain into 992 cells, most of which were concentrated in the plug, and the rest in the chamber. The width of the first cell on the back side of the plug is $2 \cdot 10^{-6}$ of the thickness of the plug. As one moves away from this cell, the sizes of cells increases such that the ratio between the width of neighbouring cells is 1.01. The initial shock in the plug elevated the thermal energy density above the rest mass density, and we run the simulation until the former drops below the latter. We tracked the hydrodynamic profiles of a single Lagrangian cell in the plug throughout the run, and plotted the hydrodynamic profiles at different times in figure \ref{fig:spherical_simulation}. The simulation confirms that the hydrodynamic variables evolve according to our theoretical predictions discussed in this section. 

It is important to point out that the values used here do not represent realistic values for any GRB scenario. This is because in a 1D simulation there can be no lateral flow, so the jet pushes all the material in front of it. Since the plug always caps the jet material, then the Lorentz factor of the jet is determined by the plug rather than the jet, which means we can neglect the particle content of the jet material.

The acceleration of a baryon shell by the jet was considered in a previous work \citep{Thompson2006DecelerationBursts}, which obtain a different scaling law for the Lorentz factor with time ($\Gamma \propto t^{1/3}$). The source of the discrepancy is a different assumption about the behaviour of photons in the jet. We assume that the jet behaves as a photon gas, which means that the photons are collisional, whereas the previous work assumes ballistic energy transport from the engine to the shell. Since the latter is less efficient, in our model the shell accelerates faster than in the previous work. We expect jets in both long and short GRBs to be in the collisional rather than the free streaming regime. We can verify this assumption by considering the optical depth for photon - photon scattering. If the energy of the photons is comparable to the rest mass energy of the electron and the cross section is comparable to the Thomson cross section, then the optical depth is

\begin{equation}
    \tau_{j \gamma \gamma} \approx \sigma_t \frac{E_j}{m_e c^2 \alpha^2 R_e^2} \approx 2 \cdot 10^{10} \frac{E_j}{10^{50} \, \rm erg} \left(\frac{R_e}{3 R_{\odot}}\right)^{-2} \left(\frac{\alpha}{0.1}\right)^{-2} \, .
\end{equation}
where $E_j$ is the energy of the jet. The large optical depth justifies our assumption that the jet is in the collisional regime. In the case of short GRBs, the optical depth will be even higher due to the smaller length scales, so the in that case the the jet will be even deeper in the collisional regime.

\section{Plug Dissolution} \label{sec:plug_dissolution}

We consider two processes that can stop the plug acceleration. The first is becoming optically thin, which we call vanishing. The second is breakup due to the Rayleigh Taylor instability.

\subsection{Vanishing}

At a sufficiently large distance the plug becomes optically thin. Beyond that point, radiative energy from the jet material can stream freely through the plug, and the acceleration stops. We consider two cases of interest. In the first we ignore pair production entirely and consider only the contribution of the original protons and electrons to the optical depth. In the second we consider the scenario where the optical depth is dominated by electron positron pairs. We call the first scenario baryon dominated, and the second scenario pair dominated.

\subsubsection{Baryon Dominated}

The plug becomes optically thin when its optical depth drops to unity
\begin{equation}
    \frac{\kappa m_{er}}{r^2} \approx 1
\end{equation}
where $\kappa$ is the opacity, which we will always assume to be Thomson opacity $\kappa \approx 0.1 cm^2/g$ and $m_{er}$ is the baryonic mass entrained in the jet eruption. We note that in the non relativistic case, photons begin to diffuse out of a shell when $\tau \approx c/v$ where $v$ is the velocity of the shell, but since we are dealing with a relativistic plug, this condition reduces to $\tau \approx 1$. In the case of long gamma ray bursts, the vanishing radius is
\begin{equation}
    r_v \approx 9 \tilde{L}^{\frac{3}{5}} \tilde{R}_{e}^{\frac{3}{5}} \tilde{\alpha}^{\frac{11}{5}} \tilde{M}_{e}^{-\frac{1}{10}} \, R_{\odot} \, .
\end{equation}
The Lorentz factor the plug attains at that point is
\begin{equation}
    \gamma_v \approx 100 \tilde{L}^{0.25} \tilde{M}_{e}^{0.06} \tilde{\alpha}^{0.14} \tilde{R}_{e}^{-0.38} \, .
\end{equation}

In the case of the short gamma ray bursts, the vanishing distance is given by
\begin{equation}
    r_v \approx 2 \tilde{L}^{\frac{1}{2}} \tilde{R}_e^{\frac{1}{2}} \tilde{\alpha}^{2} \, R_{\odot} \, .
\end{equation}
The Lorentz factor the plug attains at that point is
\begin{equation}
    \gamma_v \approx 100 \tilde{L}^{0.32} \tilde{\alpha}^{0.27} \tilde{R}_e^{-0.32} \, .
\end{equation}

\subsubsection{Pair Dominated}

In this scenario, we assume that the opacity is dominated by electron - positron pairs. This assumption introduces two changes with respect to the baryon dominated scenario. First, instead of plug mass $m_{er}$, the effective mass of the plug is given by the product of the thermal energy density and the rest volume of the plug. Second, since the effective particle mass is smaller than baryonic matter, the effective opacity is increased by the proton to electron mass ratio $m_p/m_e$. The condition for becoming optically thin in this regime is therefore
\begin{equation}
    \frac{m_p}{m_e} \frac{\kappa M}{r^2} \approx 1 \, .
\end{equation}
For the case of long gamma ray bursts, the pair dominated vanishing distance is given by
\begin{equation}
    r_v \approx 100 \tilde{L}^{0.49} \tilde{R}_{e}^{0.77} \tilde{\alpha}^{1.33} \tilde{M}_{e}^{-0.13} \, R_{\odot} \, .
\end{equation}
The Lorentz factor the plug attains at this point is
\begin{equation}
    \gamma_v \approx 600 \frac{L^{0.18} M_{e}^{0.045} \kappa^{0.23} m_{p}^{0.23}}{R_{e}^{0.27} \alpha^{0.41} c^{0.55} m_{e}^{0.23}} \, .
\end{equation}

For the case of short gamma ray bursts, the pair loaded vanishing radius is given by
\begin{equation}
    r_v \approx 180 \tilde{L}^{0.36} \tilde{R}_{e}^{0.64} \tilde{\alpha}^{1.1} R_{\odot} \, .
\end{equation}
The Lorentz factor the plug attains at this point is
\begin{equation}
    \gamma_v \approx 450 \tilde{L}^{0.23} \tilde{R}_{e}^{-0.23} \tilde{\alpha}^{-0.32} \, .
\end{equation}

Since the pair dominated plug becomes transparent at a later time, the baryon dominated regime is irrelevant. Furthermore, since breakup happens when the thermal motion in the plug is relativistic, the plug is guaranteed to be pair dominated through breakup. This is because if the particles are moving close to the speed of light, then they have enough thermal energy to produce more particles.

\subsection{Breakup} \label{sec:breakup}

The Rayleigh Taylor growth rate in the relativistic case is given by \citep{Bret2011IntuitiveRate, Matsumoto2017LinearFlow}
\begin{equation}
    \Gamma \approx \sqrt{\mathcal{A} \frac{g}{\lambda}}
\end{equation}
where $g \approx c \dot{\gamma}$ is the acceleration, $\lambda$ is the wavelength, $\mathcal{A} = \frac{h_b-h}{h_b+h}$ is the relativistic Atwood number, $h_b$ is the enthalpy of the plug and $h$ is the enthalpy of the propellant. The enthalpy is given by $h = \rho c^2 + \frac{\eta}{\eta-1} p$, where $\rho$ is the baryonic mass density. When both the propellant and the plug are radiation dominated, the pressure contribution cancels in the numerator, and since the pressure is assumed to dominate the rest mass energy density, the Atwood number in the radiative case is 
\begin{equation}
    \mathcal{A}_r \approx \frac{\rho_b c^2}{p} \, .
\end{equation}
We note that another possible scenario is that the jet is magnetically dominated, in which case propellant and the plug have different adiabatic indices ($\eta=2$ and $\eta_b=4/3$). In such a scenario the pressure contribution doesn't cancel from the numerator of the Atwood number, and so it is of order unity
\begin{equation}
    \mathcal{A}_m \approx 1 \, .
\end{equation}
In the analysis carried out in appendix \ref{sec:ped_rev}, we show that in the ultra relativistic magnetic case the plug either breaks up at the very beginning of the motion or never at all, so we will not consider this scenario, and instead focus on the ultra relativistic, hot, hydrodynamic (i.e. non magnetic) case. Magnetic fields can still be present in this picture, but their contribution to the energy density must be subdominant. In addition to that, there is another effect that comes into play when magnetic fields are present, and that is that they tend to suppress the Rayleigh Taylor instability \citep{Chandrasekhar1961HydrodynamicStability}.

We argue that the wavelength has to be comparable to the rest frame thickness of the plug $w$. This is because much larger wavelengths $\lambda \gg w$ take too long to grow, while much shorter wavelength $\lambda \ll w$ saturate before they are able to pierce the plug. After the initial shock, the plug expands adiabatically. If the plug is initially shocked to a Lorentz factor $\gamma_1$, then the shocked density $\rho_{b1}$ is greater than the unshocked density $\rho_{b0}$ by $\gamma_1$, so
\begin{equation}
    \rho_{b1} \approx \rho_{b0} \gamma_1 \, .
\end{equation}
From the conservation of rest mass (or, equivalently, conservation of baryon number)
\begin{equation}
    m \approx \alpha^2 t^2 c^2 w \rho_b \, .
\end{equation}
Using the adiabatic relation
\begin{equation}
    \rho_b \approx \rho_{b1} \left(p/p_1\right)^{1/\eta_b}
\end{equation}
where $\eta_b$ is the adiabatic index of the plug, it is possible to relate the plug rest frame thickness to the propellant pressure
\begin{equation}
    w \approx \frac{m}{\alpha^2 c^2 t^2 \rho_{b1}} \left(\frac{p}{p_1}\right)^{-1/\eta_b} \, .
\end{equation}
We also assume that the plug is initially shocked by the jet, then 
$\gamma_1 = \left(\frac{p_0}{\rho_{b0} c^2}\right)^{\left(2+\eta/\sqrt{\eta-1}\right)^{-1}}$ (see equation \ref{eq:gamma_1_planar_def}).

Breakup happens when the product of the growth rate and the rest frame time of the plug is of order unity. As is shown in \ref{sec:appendix_urm}, this condition is equivalent to the condition that the pressure in the bullet becomes comparable to the rest mass energy density. In the study of relativistic shock breakout, the latter condition determines the end of acceleration of a fluid element due to shock reflection \citep{Yalinewich2017AnalyticBreakout}. By combining the previous equations, as was described in \citep{Yalinewich2017AnalyticBreakout}, it is possible to obtain the breakup Lorentz factor
\begin{equation}
    \gamma_{br} \approx 63 \alpha^{-1.8}
\end{equation}
assuming $\eta = 4/3$. We find that this Lorentz factor only depends on the initial shocked Lorentz factor $\gamma_i$, which we assume depends only on the opening angle of the engine $\alpha$. The breakup Lorentz factor is therefore expected to be the same in both long and short GRBs.

In a similar way, it is possible to obtain the breakup radius
\begin{equation}
    r_{br} \approx r_{er} \alpha^{-1.3}
\end{equation}
where $r_{er}$ is the eruption depth. For long GRBs the breakup radius is
\begin{equation}
    r_{br,lgrb} \approx 2.5 \tilde{L}^{0.2} \tilde{R}_e^{1.2} \tilde{\alpha}^{-0.9} \tilde{M}_e^{-0.2} \, R_{\odot}\ ,.
\end{equation}
For short GRBs, the breakup radius is
\begin{equation}
    r_{br,sgrb} \approx 0.3 \tilde{h}_e \tilde{\alpha}^{-1.3} \sqrt{\ln \left(\frac{\tilde{M}_e}{\alpha^2 \tilde{L} \tilde{h}_e}\right)} R_{\odot} \, .
\end{equation}

The calculations above assume a semi infinite chamber. The fact the chamber has a finite extent means that sound waves that emanate from the plug travel inward, reflect from the centre and travel outward again. When these waves catch up to the plug, they reduce the pressure and alter its trajectory. The validity of our solution therefore depends on whether these reflected waves catch up to the plug before breakup. We saw earlier that the breakup happens at a radius larger than the stellar radius by about a factor of a few, and when the Lorentz factor is of the order of a few hundred. Even if the rarefaction wave travels at the speed of light, the radius at which it catches up to the plug would be greater than the stellar radius by the Lorentz factor squared, which is considerably larger than the breakup radius.

\section{Conclusions} \label{sec:conclusions}

In this work we developed an analytic model that describes the evolution of the baryonic material carried along when a gamma ray burst jet erupts from its envelope. For long gamma ray bursts this envelope is the stellar material, and for short gamma ray bursts it is debris from the collision of the two merging neutron stars. We find that a small amount of baryonic material (whose mass is only weakly dependent on the mass of the envelope) can be picked up by the jet, and that the interface between the stellar material and the hot jet can remain stable for a short time until it succumbs to the Rayleigh Taylor instability. During this time the stellar material is accelerated to a Lorentz factor of order $10^{1.5}$. We also show that breakup due to the Rayleigh Taylor instability occurs before the shell becomes optically thin. 

We performed some one dimensional, special relativistic hydrodynamic simulations to verify our theoretical predictions. These simulations showed that the plug evolves according to our analytic predictions. Since these were one dimensional simulations, they could only probe the evolution of the plug before the breakup. A more interesting calculation would be either a two or three dimensional simulation whose computational domain is centred around the plug, which could probe the nonlinear evolution of the Rayleigh Taylor instability. We note that Rayleigh Taylor stability of the jet was considered in previous works, but only in the lateral direction \citep{Harrison2018NumericallyMedia, Gottlieb2020TheJets}.  In those simulations, it was found that even a subdominant toroidal magnetic field can supress these instabilities  \cite{Gottlieb2020TheJets}. Simulations where the lateral instability is suppressed show a plug forming at the head of the jet that gets picked up by the jet and carried beyond the envelope.

As was mentioned earlier, previous authors showed that the existence of this baryon shell at the head of the jet can explain certain features of gamma ray bursts. However, they used different prescriptions for the trajectory of this shell, and also made different assumptions about the survival of the shell. Here we consider the implications of the results presented here to some of those models. For example, \cite{Duffell2015FromAfterglows} proposed that the accumulation of baryonic matter at the head of the jet can delay the onset of dissipation and thus give rise to late time features like gamma ray plateaus. However, their model requires the spherical cap persist up to a radius of at least 100 $R_{\odot}$, whereas we argue that the shell breaks up at much smaller radii.

We also mentioned earlier the idea that X ray flares in gamma ray bursts come from less energetic and slower moving material (relative to the jet) moving outwards from the source together with the jet. The results of our model, and in particular the plug properties we predict, favour this idea. First, we find that the plug moves at a Lorentz factor that is considerably lower than that required to circumvent the compactness problem. This can explain why emission contemporaneous with the prompt appears to an observer to be delayed, while requiring no late time activity of the source. Second, the breakup of the shell allows the jet material behind it to accelerate to a higher Lorentz factor without pushing the shell material, thus maintaining the latter's low Lorentz factor. We note that related ideas were proposed in the past to explain different properties of the prompt emission, i.e. scattering from a baryonic shell \citep{Eichler2004AnBursts, Eichler2007ABursts, Vyas2020ABursts}. However, these prompt models require a large energy discharge by the source, whereas the X ray flares release a much smaller amount of radiation, and so the energy budget considerations are less restrictive.

Finally, we would like to address some of the simplifying assumptions made in the development of our model. We assumed that the jet is stable and acts continuously. In reality, jet enginea could operator intermittently, and also wobble. In principle, both of these effects could invalidate all the results derived in this work. However, if these are taken to the extreme, e.g. if the duty cycle is too narrow or if the wobble angle is too large, then the jet might fail to break out of the star altogether \citep{Gottlieb2020IntermittentGRBs}. We note that it has been suggested that with a mild magnetisation a jet can withstand these distortions and still break out of a star \citep{Gottlieb2021IntermittentGRBs}. It would be interesting to explore the parameter space in which these effects change the behaviour of the plug while still allowing for a successful jet breakout. The answers to these questions are beyond the scope of this model, and we relegate them to a future work.

\section*{Acknowledgements}

AY would like to thank Chris Thompson, Ehud Nakar and Ore Gottlieb for the useful discussions. AY is supported by the Natural Sciences and Engineering Research Council of Canada (NSERC), funding reference \#CITA 490888-16. PB's research was supported by a grant (no. 2020747) from the United States-Israel Binational Science Foundation (BSF), Jerusalem, Israel. This work made use of the numpy \citep{Harris2020ArrayNumPy}, sympy \citep{Meurer2017SymPy:Python} and matplotlib \citep{Hunter2007Matplotlib:Environment} python packages. 

\section*{Data Availability}

The source code for the numerical simulation and documentation can be found on github at \url{https://github.com/bolverk/fujin}. 




\bibliographystyle{mnras}
\bibliography{references} 




\appendix

\section{Pedagogical Review} \label{sec:ped_rev}

In this section we present some toy models that demonstrate the principles used to perform the calculations described in the body of the paper.

\subsection{Newtonian Motion}

Let us consider a plug of mass $m$ and density $\rho_b$ in a barrel with cross section $A$, being pushed by a propellant gas with pressure $p$ and density $\rho \ll \rho_b$. The pressure behind the plug is roughly constant until the plug is accelerated to a velocity comparable to the propellant speed of sound $a = \sqrt{\eta p/\rho}$, where $\eta$ is the propellant adiabatic index. The time it takes for the plug to attain a velocity comparable to the propellant speed of sound is
\begin{equation}
    \Delta t \approx \frac{m a}{A p} \, .
\end{equation}
Since the plug is denser than the propellant, the interface between them is Rayleigh Taylor unstable. The Rayleigh Taylor growth rate is given by
\begin{equation}
    \Gamma \approx \sqrt{\frac{g}{\lambda}}
\end{equation}
where $g \approx p A/m$ is the acceleration and $\lambda$ is the wavelength of the perturbation. We argue that the relevant range of wavelengths has to be comparable to the thickness of the plug $w = m/A \rho_b$. This is because much larger wavelengths $\lambda \gg w$ take too long to develop, and much smaller wavelengths $\lambda \ll w$ develop fast, but also saturate fast. Hence, only modes with $\lambda \approx w$ can break up the plug. For this to happen, the product of the growth rate and time (which we call growth) should be greater than unity. In our case
\begin{equation}
    G = \Gamma \Delta t \approx \sqrt{\frac{\rho_b}{\rho}} \, .
\end{equation}
We find that in the Newtonian case the plug would break immediately.

For completeness we also derive the trajectory of a Newtonian plug. The equation of motion for the plug is given by
\begin{equation}
    m \frac{d v}{d t} = A p \, .
\end{equation}
Each displacement of the projectile sends out a sonic wave that travels backwards in the propellant. The changes in velocity and pressure are therefore related through
\begin{equation}
    dv = \frac{d p}{a \rho} \, .
\end{equation}
The speed of sound $a$ and density $\rho$ of the propellant can be related to the pressure using the isentropic relations
\begin{equation}
    \rho = \rho_1 \left(p/p_1\right)^{1/\eta}
\end{equation}
and
\begin{equation}
    a = a_1 \left(p/p_1\right)^{\frac{\eta-1}{2 \eta}}
\end{equation}
where $p_1$, $\rho_1$ and $a_1$ are the pressure, density and speed of sound at the beginning of the motion. With these substitutions we can solve the equation of motion for the pressure
\begin{equation}
    p = p_1 \left(1+\frac{\eta+1}{2 \eta} \frac{A a_1 \rho_1}{m} t\right)^{-\frac{2 \eta}{\eta+1}} \, .
\end{equation}
The velocity is given by
\begin{equation}
    v  = \frac{2 a_1}{\eta-1} \left[1-\left(1+\frac{\eta+1}{2 \eta} \frac{A a_1 \rho_1}{m} t \right)^{-\frac{\eta-1}{\eta+1}}\right] \, .
\end{equation}

\subsection{Ultra Relativistic Motion} \label{sec:appendix_urm}

The major difference between the newtonian case and the ultra relativistic case is that in the latter the speed of sound is comparable to the speed of light, and remains so until the propellant pressure becomes comparable with the propellant rest mass energy density. We distinguish between two different limits in this regime. In the first, which we refer to as ``cold" $p \ll \rho_b c^2$, and in the other, which we refer to as ``hot" $p \gg \rho_b c^2$.

\subsubsection{Cold Case} \label{sec:apndx_rel_cold}

The growth rate in this case is the same as in the non relativistic case. This is because one could always consider the problem from a co-moving reference frame, where the motion of the plug is non relativistic. In this frame, the time interval in which the plug remains non relativistic is given by
\begin{equation}
    \Delta t' \approx \frac{m c}{A p} \, .
\end{equation}
The net growth is
\begin{equation}
    G = \Delta t' \Gamma \approx \sqrt{\frac{\rho_{b0} c^2}{p}} \label{eq:rel_cold}
\end{equation}
where $\rho_{b0}$ is the initial density of the plug.
Hence in this limit the behaviour is the same as in the Newtonian case and the plug breaks up immediately.

\subsubsection{Hot Case}

In the previous cases we assumed that the mass density in the plug is much larger than the mass density in the propellant, but now this assumption is no longer valid. This is because now the thermal pressure $p$ is much larger than the rest mass energy of the plug. This means that the energy densities in the plug and in the target are at least comparable. This effect reduces the Rayleigh Taylor growth rate by the so called Atwood number $\mathcal{A} = \left(h_b - h\right)/\left(h_b+h\right)$ where $h_b$ is the enthalpy of the plug and $h$ is the enthalpy of the target. The full expression for the growth rate is given by
\begin{equation}
    \Gamma = \sqrt{\mathcal{A} g / \lambda} \, .
\end{equation}
If both have different adiabatic indices, then then the Atwood number is of order unity. The effective mass of the plug is $m \approx A w p/c^2$, and so the growth is 
\begin{equation}
    G \approx \Delta t' \Gamma \approx 1 \, .
\end{equation}
Hence in this case the plug also breaks up immediately.

If both adiabatic indices are comparable, then the enthalpies are almost the same, and the difference between them is comparable to the plug rest mass energy density \citep{Bret2011IntuitiveRate}. The Atwood number in this case is very small, and is given by
\begin{equation}
    \mathcal{A} \approx \frac{\rho_b c^2}{p}
\end{equation}
the growth rate is
\begin{equation}
    \Gamma \approx \sqrt{\frac{\rho_b c^2 A}{m w}}
\end{equation}
and the growth is
\begin{equation}
    G \approx \sqrt{\frac{\rho_b c^2}{p}} \ll 1 \, .
\end{equation}
Finally, we find a system that behaves in an interesting way. A plug in this regime does not disintegrate until the pressure becomes comparable with the plug rest mass energy density. Since the plug is expanding adiabatically, the pressure can be obtained by solving $p/p_1 = \left(p/\rho_{1b} c^2\right)^{\eta_b}$, and is given by
\begin{equation}
    p = p_1 \left(\frac{\rho_{b1} c^2}{p_1}\right)^{\frac{\eta_b}{\eta_b-1}}
\end{equation}
where $p_1$ is the the propellant pressure in the beginning of the motion and $\rho_{b1}$ is plug mass density at the beginning of the motion.

\begin{figure*}
\includegraphics[width=0.9\textwidth]{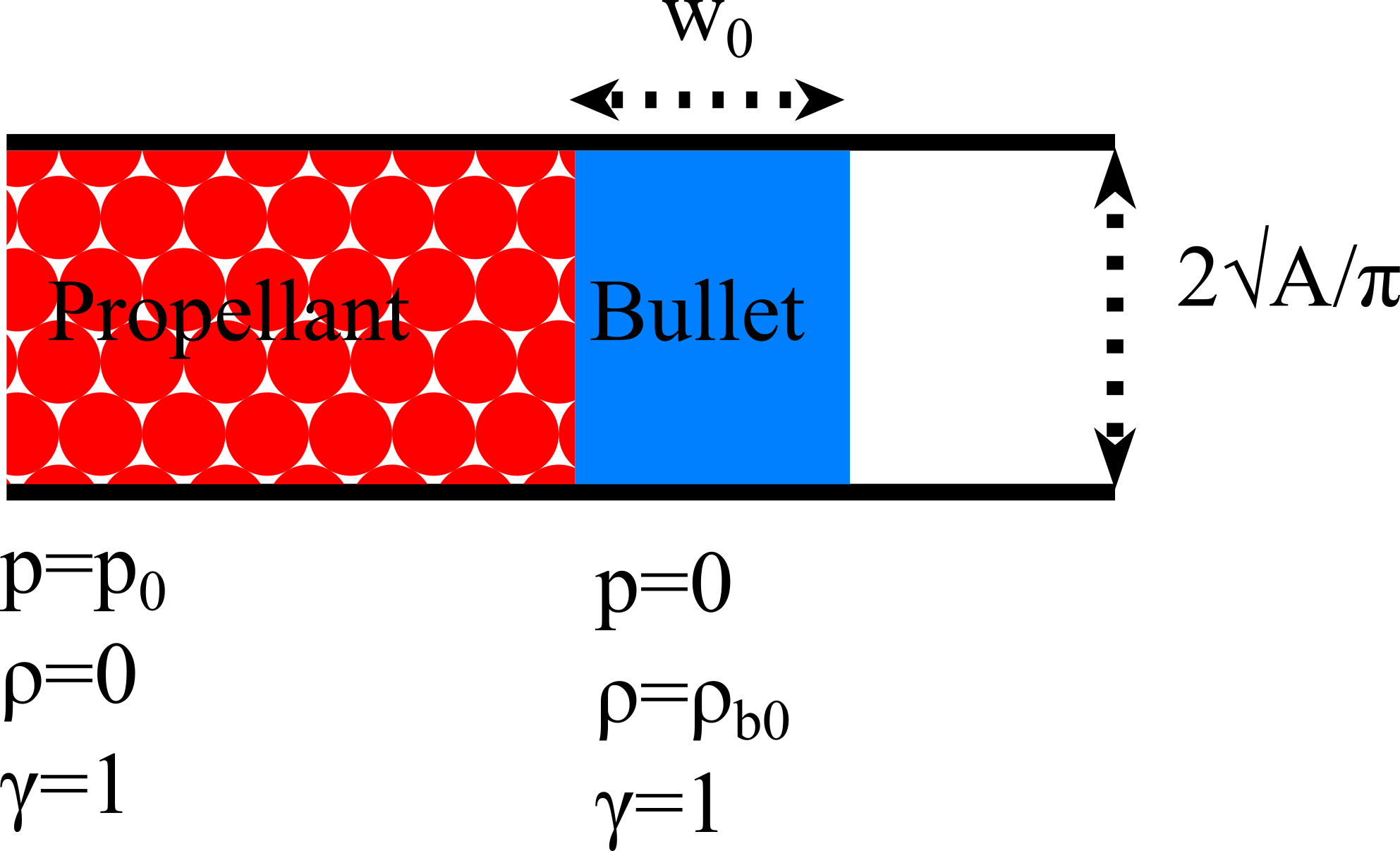}
\caption{A schematic illustration of the plug and the propellant in the barrel before the first shock.}
\label{fig:planar_schematic_0}
\end{figure*}

\begin{figure*}
\includegraphics[width=0.9\textwidth]{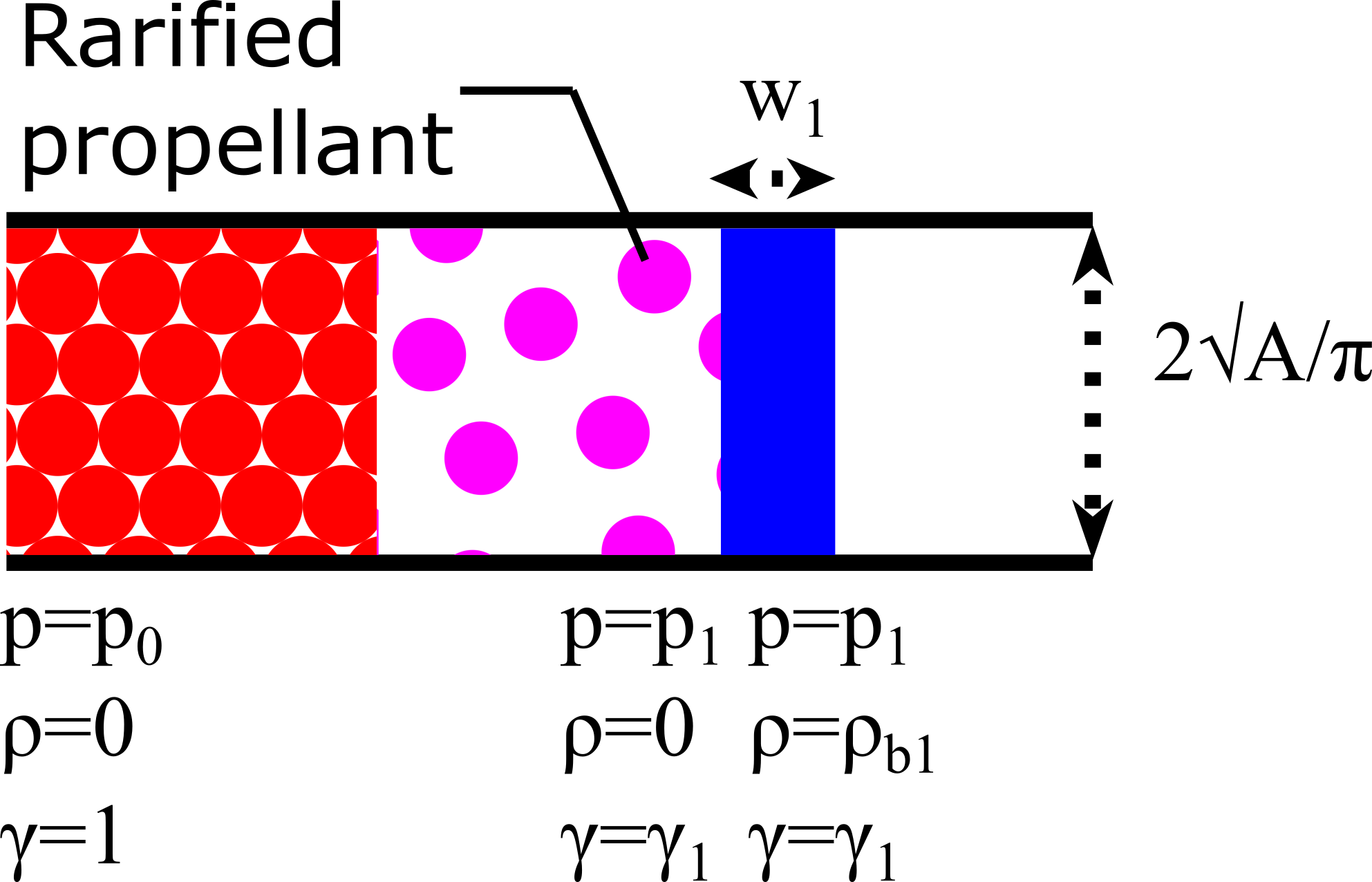}
\caption{A schematic illustration of the plug and propellant in the barrel after the passage of the first shock through the plug.}
\label{fig:planar_schematic_1}
\end{figure*}

\begin{figure*}
    \centering
    \includegraphics[width=0.9\textwidth]{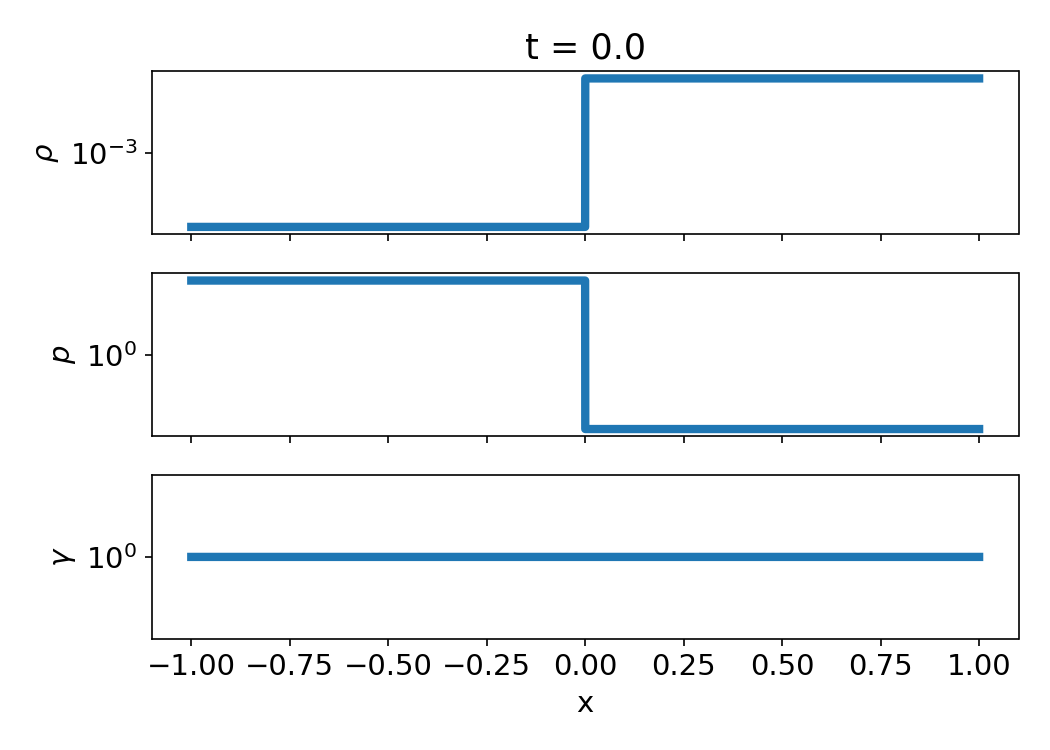}
    \caption{Initial conditions for the simulation of the wave trajectories. The propellant with high pressure and low density is on the left, and the plug with high density and low pressure is on the right.}
    \label{fig:rp0}
\end{figure*}

\begin{figure*}
    \centering
    \includegraphics[width=0.9\textwidth]{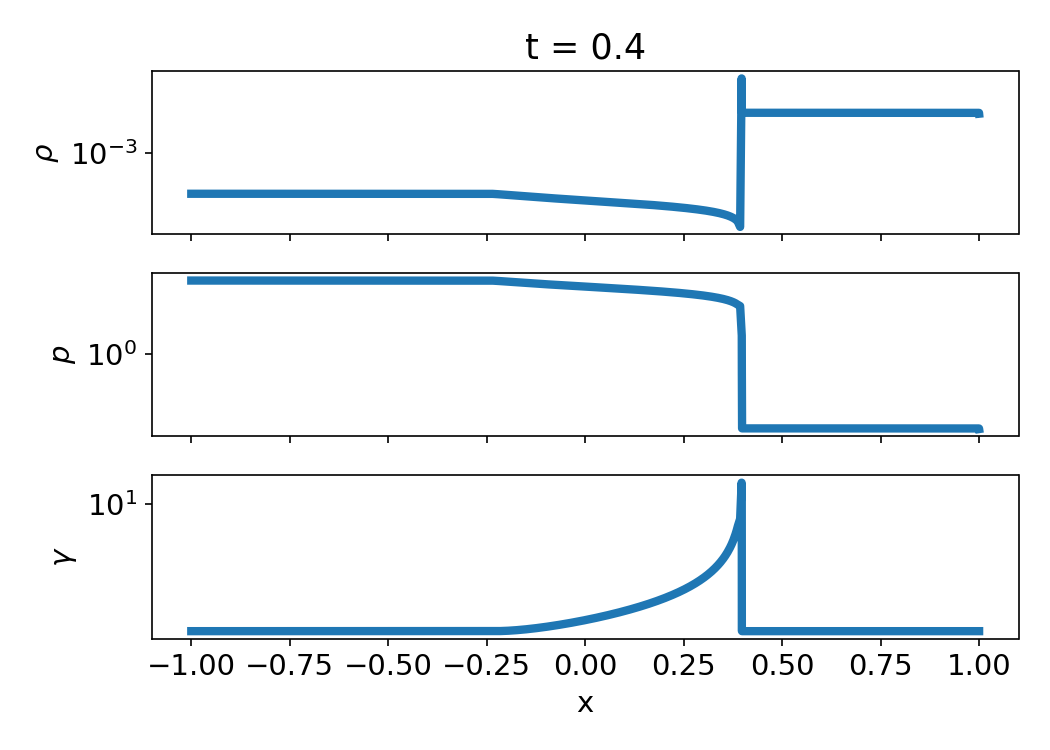}
    \caption{Final conditions for the simulation for the wave trajectories. A shock wave can be seen moving to the right, and a rarefaction wave moving to the left.}
    \label{fig:rp1}
\end{figure*}

To find the time when the pressure becomes comparable to the rest mass energy, we need to obtain the trajectory. The equation of motion for the plug is
\begin{equation}
    \frac{d}{d t} \left(M \gamma \beta\right) = \frac{A p}{c}
\end{equation}
where $M$ is the inertia of the plug in its rest frame. This inertia is proportional to the product of the thickness of the plug in the rest frame and the pressure $M \propto w' p$, so if the plug expands adiabatically, the inertia scales with the pressure as $M \propto p^{1-1/\eta_b}$. To close the equation of motion, we need another relation between the velocity and pressure. In analogy to the non relativistic case, this relation is provided by the conservation of the ultra relativistic Riemann invariant \citep{Johnson1971RelativisticDimension}
\begin{equation}
    J_+ = p \gamma^{\frac{\eta}{\sqrt{\eta-1}}} \, . \label{eq:urri}
\end{equation}
With equation \ref{eq:urri} we can solve the equation of motion to obtain the pressure and Lorentz factor as a function of time
\begin{equation}
    \gamma = \gamma_1 \left(\frac{A p_{1} t}{M_1 c \gamma_1}\right)^{\frac{\eta_{b} \sqrt{\eta - 1}}{\eta + \eta_{b} \sqrt{\eta - 1}}}
\end{equation}
\begin{equation}
    p = p_1 \left(\frac{M_1 c \gamma_{1}}{A t p_1}\right)^{\frac{\eta \eta_{b}}{\eta + \eta_{b} \sqrt{\eta - 1}}}
\end{equation}
Where $p_1$, $\gamma_1$ and $M_1$ are the initial pressure, Lorentz factor and rest frame inertia of the plug. In the scenario we have in mind, before the plug began to move it was perfectly cold, with mass density $\rho_{b0}$ and thickness $w_0$, and the propellant had a pressure $p_0$. As a result of the first contact between the hot propellant and the plug, a shock wave emerges from the contact and moves into the plug, and a rarefaction wave moves into the propellant. The wave trajectories are demonstrated schematically in figures \ref{fig:planar_schematic_0} and \ref{fig:planar_schematic_1} and using  a relativistic hydrodynamic simulation in figures \ref{fig:rp0} and \ref{fig:rp1}. The post shock pressure and Lorentz factor can be found by solving the Riemann problem on the interface. On the shock $p_1 \approx \gamma_1^2 \rho_{b0} c^2$, and on the rarefaction wave $p_1 = p_0 \gamma_1^{-\frac{\eta}{\sqrt{\eta-1}}}$. The solution is
\begin{equation}
    \gamma_1 = \left(\frac{p_0}{\rho_{b0} c^2}\right)^{\left(2+\eta/\sqrt{\eta-1}\right)^{-1}}
    \label{eq:gamma_1_planar_def}
\end{equation}
\begin{equation}
    p_1 = p_0 \left(\frac{p_0}{\rho_{b0} c^2}\right)^{-\left(2 \sqrt{\eta-1}/\eta + 1\right)^{-1}} \, .
\end{equation}
The post shock rest mass density in the plug is
\begin{equation}
    \rho_{b1} = \rho_{b0} \gamma_1 = \rho_{b0} \left(\frac{p_0}{\rho_{b0} c^2}\right)^{\left(2+\eta/\sqrt{\eta-1}\right)^{-1}} \, .
\end{equation}
Due to the conservation of baryon number, the post shock rest frame width of the plug is $w_1 = w_0 \rho_{b0}/\rho_{b1}$. The initial inertial mass is give by
\begin{equation}
    M_1 = A w_1 p_1/c^2 = A w_0 \rho_{b0} \left(\frac{p_0}{\rho_{b0} c^2}\right)^{\left(2+\eta/\sqrt{\eta-1}\right)^{-1}} \, .
\end{equation}

Now we can find the time at which the plug disintegrates. To simplify the expression, we assume that $\eta_b = \eta$. The break time in the lab frame (where the plug was stationary before the motion started) is given by
\begin{equation}
    t_b = \frac{M_1 c \gamma_1}{A p_1} \left(\frac{p_1}{\rho_{b1} c^2}\right)^{\frac{1+\sqrt{\eta-1}}{\eta-1}} = \frac{w_0}{c} \left(\frac{p_0}{\rho_{b0} c^2}\right)^{\left(1+\sqrt{\eta-1}\right)^{-1}} \, .
\end{equation}

\begin{figure*}
    \centering
    \includegraphics[width=0.9\textwidth]{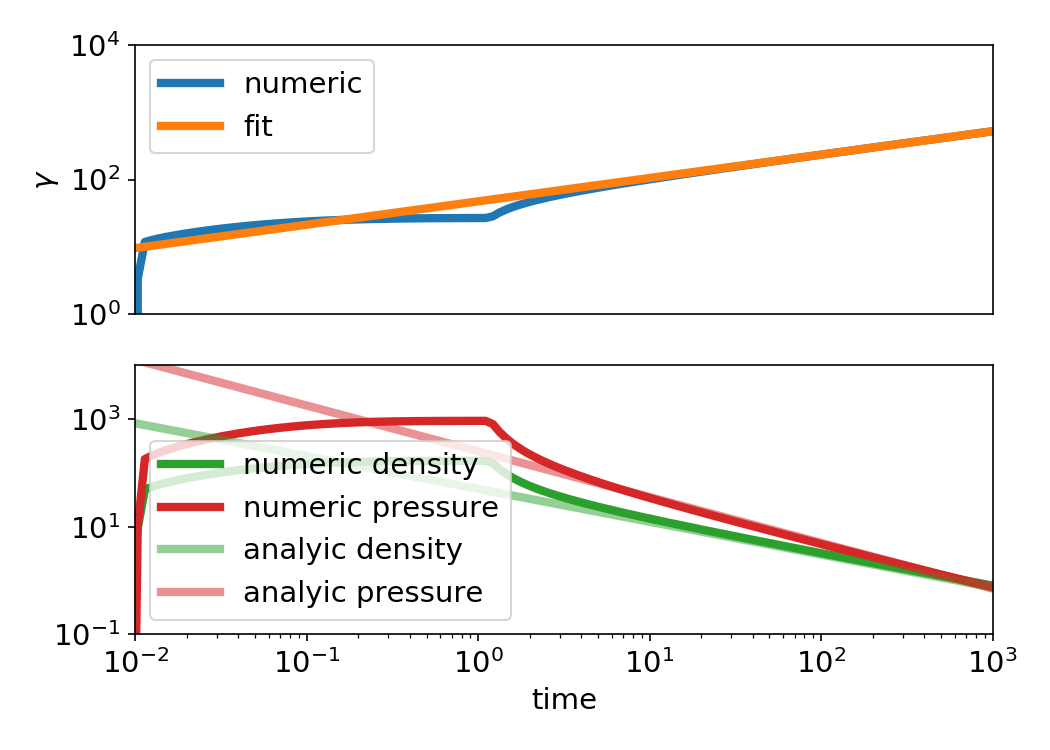}
    \caption{Histories of hydrodynamic values of a single fluid element inside the plug, in the case of an acceleration inside a planar barrel. The numerical slope of the Lorentz factor history is $d \ln \gamma / d \ln t = 0.35$, which is very close to the theoretical value 0.37. The bottom panel shows the evolution of the pressure and density. The thermal pressure initially exceeds the rest mass energy density, but as the plug cools and expands the pressure eventually drops below the rest mass energy density.}
    \label{fig:planar_simulation}
\end{figure*}

To verify our results, we ran a one dimensional, Lagrangian, special relativistic hydrodynamic simulation. The initial computational domain lay between $0<x<10$, where $0<x<9$ represents the chamber and $9<x<10$ represents the plug, both of which are described by an ideal gas equation of state with an adiabatic index $\eta=4/3$. Inside the chamber the density was low ($10^{-8}$) and the pressure high ($10^7$) while in the plug the density was high (1) and the pressure low ($10^{-8}$). The velocity was zero everywhere. We divided the computational domain into 890 cells, most of which were concentrated in the plug, and the rest in the chamber. The initial shock in the plug elevated the thermal energy density above the rest mass density, and we run the simulation until the former drops below the latter. We tracked the hydrodynamic profiles of a single Lagrangian cell in the plug throughout the run, and plotted the hydrodynamic profiles at different times in figure \ref{fig:planar_simulation}. The simulation confirms that the hydrodynamic variables evolve according to our theoretical predictions discussed in this section.

\subsection{Spherical Relativistic Riemann Invariant} \label{sec:spheric_ri}

We begin this derivation with the better known planar Riemann invariant \citep{Johnson1971RelativisticDimension}
\begin{equation}
J^{p}_{\pm} = p \gamma^{\pm \eta/\sqrt{\eta-1}} \, . \label{eq:planar_ri}
\end{equation}
Using the slab symmetric one dimensional special relativistic hydrodynamic equations
\begin{equation}
    \frac{1}{c}\frac{\partial}{\partial t} \left(\left(e+\beta^2 p\right) \gamma^2\right) + \frac{\partial}{\partial r} \left(\left(e+p\right) \beta \gamma^2\right) = 0
\end{equation}
and
\begin{equation}
    \frac{1}{c}\frac{\partial}{\partial t} \left(\left(e+p\right) \beta \gamma^2\right) + \frac{\partial}{\partial r} \left(\left(e \beta^2 +p\right) \gamma^2\right) = 0
\end{equation}
where $e = p/\left(\eta-1\right)$, it can be verified that the planar relativistic Riemann invariant (equation \ref{eq:planar_ri}) is conserved along flow line, i.e.
\begin{equation}
    \left(\frac{1}{c}\frac{\partial}{\partial t} + \frac{\beta \pm \beta_a}{1\pm \beta_a \beta}\right) \log J^p_{\pm} = 0 \label{eq:characteristic_equation}
\end{equation}
where $\beta_a = \sqrt{\eta-1}$ is the speed of sound of a relativistic fluid. We note that in the planar case there are two conserved quantities, one for forward going waves and another for backward going waves. In the remainder of this derivation we will only treat the forward Riemann invariant.

In spherically symmetric flow, the hydrodynamic equations take the form
\begin{equation}
    \frac{1}{c}\frac{\partial}{\partial t} \left(\left(e+\beta^2 p\right) \gamma^2\right) + \frac{1}{r^2}\frac{\partial}{\partial r} r^2 \left(\left(e+p\right) \beta \gamma^2\right) = 0
\end{equation}
and
\begin{equation}
    \frac{1}{c}\frac{\partial}{\partial t} \left(\left(e+p\right) \beta \gamma^2\right) + \frac{1}{r^2} \frac{\partial}{\partial r} \left(r^2 \left(e + p\right) \beta^2 \gamma^2\right) + \frac{\partial p}{\partial r} = 0 \, .
\end{equation}
If we do substitute the planar Riemann invariant into the left hand side of equation \ref{eq:characteristic_equation}, and use the spherical hydrodynamic equations, we find that this quantity is no longer conserved on characteristics
\begin{equation}
    \left(\frac{1}{c}\frac{\partial}{\partial t}+\frac{\beta_a+\beta}{1+\beta_a \beta}\right) \log J^p_+ = - \frac{2\sqrt{\eta-1}}{r \left(\sqrt{\eta-1}+1\right)} \, .
\end{equation}
Since we are interested in ultra - relativistic flows, we can replace the radius on the right hand side with time $r = c t$. With this new substitution, we can incorporate the extra term into the planar Riemann invariant and obtain a modified, spherical Riemann invariant that is conserved in spherically symmetric flow
\begin{equation}
    J_+ = p \gamma^{\frac{\eta}{\sqrt{\eta-1}}} t^{\frac{2 \eta}{\sqrt{\eta-1}+1}} \, .
\end{equation}


\bsp	
\label{lastpage}
\end{document}